\documentclass[
aip,
 amsmath,amssymb,
 reprint,%
]{revtex4-1}

\usepackage{graphicx}
\usepackage{dcolumn}
\usepackage{bm}

\usepackage[utf8]{inputenc}
\usepackage[T1]{fontenc}
\usepackage{mathptmx}
\usepackage{etoolbox}
\usepackage{hyperref}

\makeatletter
\def\@email#1#2{%
 \endgroup
 \patchcmd{\titleblock@produce}
  {\frontmatter@RRAPformat}
  {\frontmatter@RRAPformat{\produce@RRAP{*#1\href{mailto:#2}{#2}}}\frontmatter@RRAPformat}
  {}{}
}%
\makeatother

\begin{document}

\preprint{AIP/123-QED}

\title[Symbolic dynamics of joint brain states during dyadic coordination]
{Symbolic dynamics of joint brain states during dyadic coordination}

\author{Italo Ivo Lima Dias Pinto}
\affiliation{Department of Cognitive Sciences, University of California, Irvine, CA 92617, U.S.A.}
\affiliation{U.S. Army DEVCOM Army Research Laboratory, Humans in Complex Systems Division, Aberdeen Proving Ground, MD 21005, U.S.A.}
\email{ilimadia@uci.edu}
\author{Zhibin Zhou}
\affiliation{Department of Cognitive Sciences, University of California, Irvine, CA 92617, U.S.A.}

\author{Javier O. Garcia}
\affiliation{U.S. Army DEVCOM Army Research Laboratory, Humans in Complex Systems Division, Aberdeen Proving Ground, MD 21005, U.S.A.}

\author{Ramesh Srinivasan}
\affiliation{Department of Cognitive Sciences, University of California, Irvine, CA 92617, U.S.A.}
\affiliation{U.S. Army DEVCOM Army Research Laboratory, Humans in Complex Systems Division, Aberdeen Proving Ground, MD 21005, U.S.A.}
\affiliation{Department of Biomedical Engineering, University of California, Irvine, CA 92617, U.S.A.}


\date{\today}
\begin{abstract}
We propose a novel approach to investigate the brain mechanisms that support coordination of behavior between individuals. Brain states in single individuals defined by the patterns of functional connectivity between brain regions are used to create joint symbolic representations of brain states in two or more individuals to investigate symbolic dynamics that are related to interactive behaviors.  We apply this approach to electroencephalographic (EEG) data from pairs of subjects engaged in two different modes of finger-tapping coordination tasks (synchronization and syncopation) under different interaction conditions (Uncoupled, Leader-Follower, and Mutual) to explore the neural mechanisms of multi-person motor coordination. Our results reveal that dyads exhibit mostly the same joint symbols in different interaction conditions - the most important differences are reflected in the symbolic dynamics.  Recurrence analysis shows that interaction influences the dwell time in specific joint symbols and the structure of joint symbol sequences (motif length). In synchronization, increasing feedback promotes stability with longer dwell times and motif length.  In syncopation, Leader-Follower interactions enhance stability (increase dwell time and motif length), but Mutual interaction dramatically reduces stability. Network analysis reveals distinct topological changes with task and feedback. In synchronization, stronger coupling stabilizes a few states, preserving a core-periphery structure of the joint brain states while in syncopation we observe a more distributed flow amongst a larger set of joint brain states. This study introduces symbolic representations of metastable joint brain states and associated analytic tools that have the potential to expand our understanding of brain dynamics in human interaction and coordination. 

\end{abstract}

\maketitle


\begin{quotation}

Experimental studies of the neural mechanisms of coordination between individuals is an emerging field of neuroscience.  This study proposes a new approach to model multi-brain dynamics by introducing symbolic representations of metastable joint brain states. Symbolic dynamics of joint brain states can be characterized by recurrence analysis and transition networks to distinguish motor coordination tasks performed with different interaction modes. 

\end{quotation}

\section{Introduction}

Coordination and interaction between humans are an essential cognitive function. There has been growing interest in understanding the brain mechanisms that support such social cognition by simultaneous recording of brain activity during interactions between multiple individuals, an experimental procedure known as hyperscanning \cite{montague2002hyperscanning}. Inspired by the development of symbolic dynamics and its application to brain activity \cite{graben2013detecting,beim2019metastable,hutt2017sequences,de2021symbolic,da2024symbolic}, we propose a new approach to modeling the relationship between brain activity in two or more individuals as \emph{joint symbolic dynamics}. The symbol sequences for each individual are defined by brain microstates \cite{michel2018eeg,lehmann1987eeg,lehmann2009eeg,allefeld2009mental}, estimated from patterns of functional connectivity.  The  joint symbol sequences provide an entry point to investigate the metastable dynamics of brain states that support coordination in behavior.  

Experimental studies of the neural mechanisms of coordination between individuals is an emerging field both at behavioral level \cite{guevara2017attractor,haken1985theoretical} and in the brain activity level, particularly in the context of coordination of movement \cite{moreau2023performance} and other aspects of social cognition \cite{babiloni2014social}. One approach in these studies is to simultaneously measures of electroencephalography (EEG) in two or more participants to study the brain dynamics associated with the coordination of the behavior.  Although these have been innovative experiments, previous analysis has simply extended the conceptual framework of interactions between brain regions within a single brain, typically through measures of correlations or synchrony, to multiple brains.  Measures like Partial Directed Coherence\cite{babiloni2006hypermethods,wehling2008assessment} and Phase Locking Value\cite{dumas2010inter} are used to assess synchrony between different brain areas of different subjects\cite{babiloni2014social}. Within a single brain, axonal connections (white-matter) are the physiological basis for synchronization between brain regions\cite{nunez2014neocortical,allefeld2004approach}.  Across brains, there is no physiological basis for direct neural interactions on the millisecond time scale. Synchrony across brains likely largely results from common sensory inputs \cite{hasson2004intersubject} or even correlations observed in simultaneous EEG recordings that occur due to unknown mechanisms \cite{wackermann2003correlations}. Typically, studies that measure synchrony across brains find very small (typically correlations < 0.01) levels of synchrony, which can often lead to challenges in interpretation\cite{burgess2013interpretation}  


Brain activity measured at a macroscopic (cm) scale in humans using electroencephalography (EEG) reflects transient quasi-stable patterns that evolve over time\cite{nunez2006electric}. An extensive literature characterizes these patterns as functional networks, using correlation, coherence, or mutual information to identify structure in patterns of connectivity between brain regions.  One approach to EEG analysis is to identify stable patterns in short windows of time, called microstates, and to characterize the EEG signals as a sequence of microstates \cite{michel2018eeg}. By labeling each microstate with a discrete symbol, various approaches based on information theory have been applied to the symbolic dynamics to characterize the evolution of brain states \cite{hutt2017sequences}.

Inspired by the symbolic dynamics of brain microstates, we have developed an approach to multi-brain dynamics to investigate transitions between joint metastable states by representing functional brain networks as multi-brain symbolic sequences. From a dynamical systems perspective, it is understood that metastability can emerge from perturbations, whether they are stochastic in nature\cite{pinto2014globally,rosas2016globally} or caused by external forcing. A prime example of such phenomena is stochastic resonance \cite{gammaitoni1998stochastic}, in which both stochastic and external periodic forcing play a role in inducing periodic transitions between metastable states. This approach may be particularly fruitful in investigating coordination of behavior where two stochastic systems are coupled.   


Using a data-driven approach, we define brain states based on the correlations between different brain areas for each subject, allowing us to create symbolic representations of joint brain states. This paradigm enables the exploration of multi-brain dynamics without relying on millisecond time scale connectivity between brains. In our analysis we studied multi-brain symbolic dynamics using Recurrence Quantification Analysis (RQA), which provide powerful tools to analyze and understand the intricate dynamical processes governing metastable states in the brain\cite{hutt2017sequences,beim2019metastable,graben2013detecting}. Additionally, we investigate the symbolic dynamics by constructing transition networks\cite{zou2019complex,masoller2015quantifying}, providing deeper insights into the topology and dynamics of interconnected brain activity.

The paper is organized as follows: In section \ref{Symbolic_single} we describe the Machine Learning pipeline used to determine a single subject's symbolic sequence. Section \ref{Toy_model} we apply the methodology used to determine the symbolic sequence on a well studied toy model as a proof of concept. Section \ref{Symbolic_joint} we extend the notion of symbolic sequence to dyads and present summary statistics of the symbol sequences observed in an experiment with motor coordination under different interaction conditions. Section \ref{RQA} presents a recurrent quantification analysis of the symbolic sequence of the joint states. Section \ref{Topology} present the joint symbolic sequence as a transition network, to investigate the topology of the brain states represented in the symbol sequence. In section \ref{Discussion} we discuss and interpret our results. Section \ref{Conclusion} summarizes our main findings and the potential of our approach.

\section{Symbolic representation of brain states defined by functional connectivity}
\label{Symbolic_single}

\begin{figure*}[htb]
\centering
\includegraphics[width=0.9\textwidth]{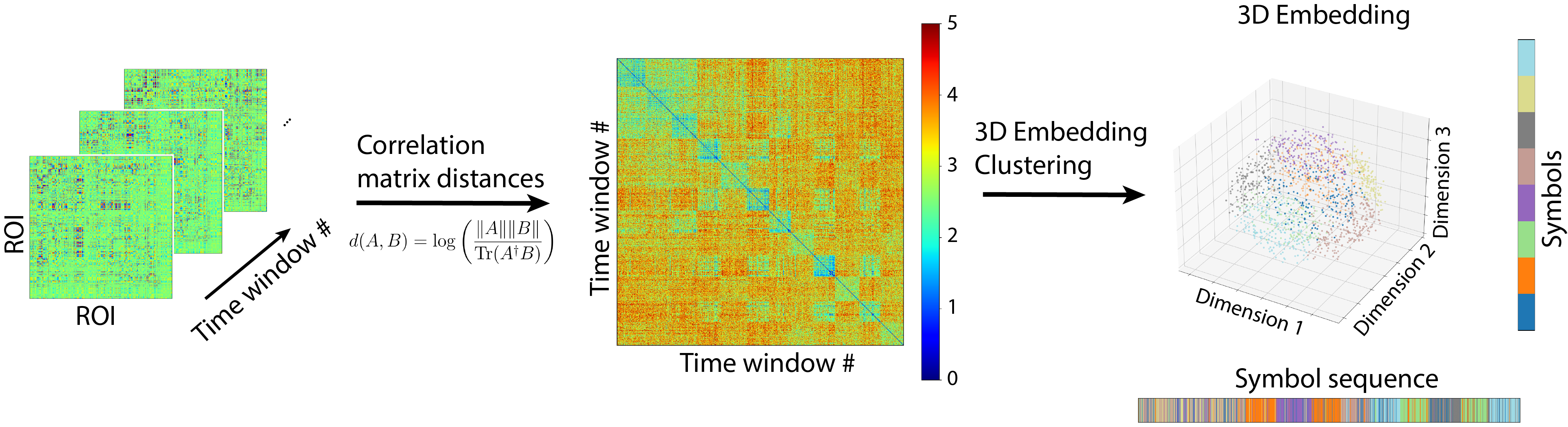}
\caption{Symbolic representation of brain states. Brain states were characterized by measuring the strength of functional connectivity between all pairs of EEG signals, represented as a unsigned correlation matrix.  The figure on the left shows a time series of correlation matrices.  The dissimilarity of the correlation matrices was characterized by a distance measure (Eq. \ref{eqn_dist}) and a distance matrix was calculated as shown in the middle figure.  Multidimensional scaling was used to define a low-dimensional embedding of the correlation matrices approximating the distance matrix and a Gaussian mixture model was used to cluster the correlation matrices, and identify discrete brain states labeled as symbols, color coded on the right figure. Below the figure on the right is an example of a symbol sequence representing the multidimensional EEG time series}
\label{fig1}
\end{figure*}

In this study we characterized brain states by the pattern of functional connectivity observed in short epochs of EEG signals (2s long with 50\% overlap). We obtained an analytic signal using a Hilbert Transform, and for each epoch computed the complex-valued correlation between each pair of analytic signals to obtain a time series of correlation matrices, as show in Fig. \ref{fig1}. This approach allows the detection of delays between signals due to axonal transmission\cite{nunez2014neocortical}. These correlation matrices are used to identify a discrete set of brain states for each subject which are the \emph{symbols} used to represent the time series.  The distance $d$ between correlation matrices $A$ and $B$ can be defined as:
\begin{equation}
d = \log \left(\frac{\|A\| \|B\|}{\text{tr}(A^\dagger B)}\right),
\label{eqn_dist}
\end{equation}
where $A^\dagger = \left( \overline{A} \right)^T$. 
If $ A $ and $ B $ are highly aligned, $ \text{tr}(A^\dagger B) $ approaches $\|A\| \times \|B\|$, making the fraction close to 1 and thus the logarithm close to 0, indicating low distance or high similarity. This metric is a modification of an existing metric for correlation matrices \cite{herdin2005correlation}, however in the original form the distance has a upper bound of 1, while in our metric there is no upper bound increasing its sensitivity for correlation matrices with larger distances.

The distance is calculated for each pair of correlation matrices in a given session resulting in a distance matrix for each subject across an entire experimental session, as shown in Fig. \ref{fig1}. Multidimensional scaling (Scikit-learn sklearn.manifold.MDS class \cite{borg2007modern}) was used to create a 3 dimensional embedding (dimensionality selected using Bayesian Information Criterion) of the correlation matrices as shown in Fig. \ref{fig1}.  Discrete brain states were then identified by fitting a Gaussian mixture model (Scikit-learn sklearn.mixture.GaussianMixture class \cite{reynolds2009gaussian}), selecting the number of clusters that minimize Bayesian Information Criteria. Each of those coarse grained states were labeled with \emph{symbols}. The term \emph{symbolic sequence} refers to the time sequence of brain states and the term \emph{symbol} is used to refer to the brain states in those sequences. A schematic representation of the sequence derived for a single experiment is shown in Fig. \ref{fig1}.

\section{State detection validation}
\label{Toy_model}

\begin{figure*}[htb]
\centering
\includegraphics[width=0.9\textwidth]{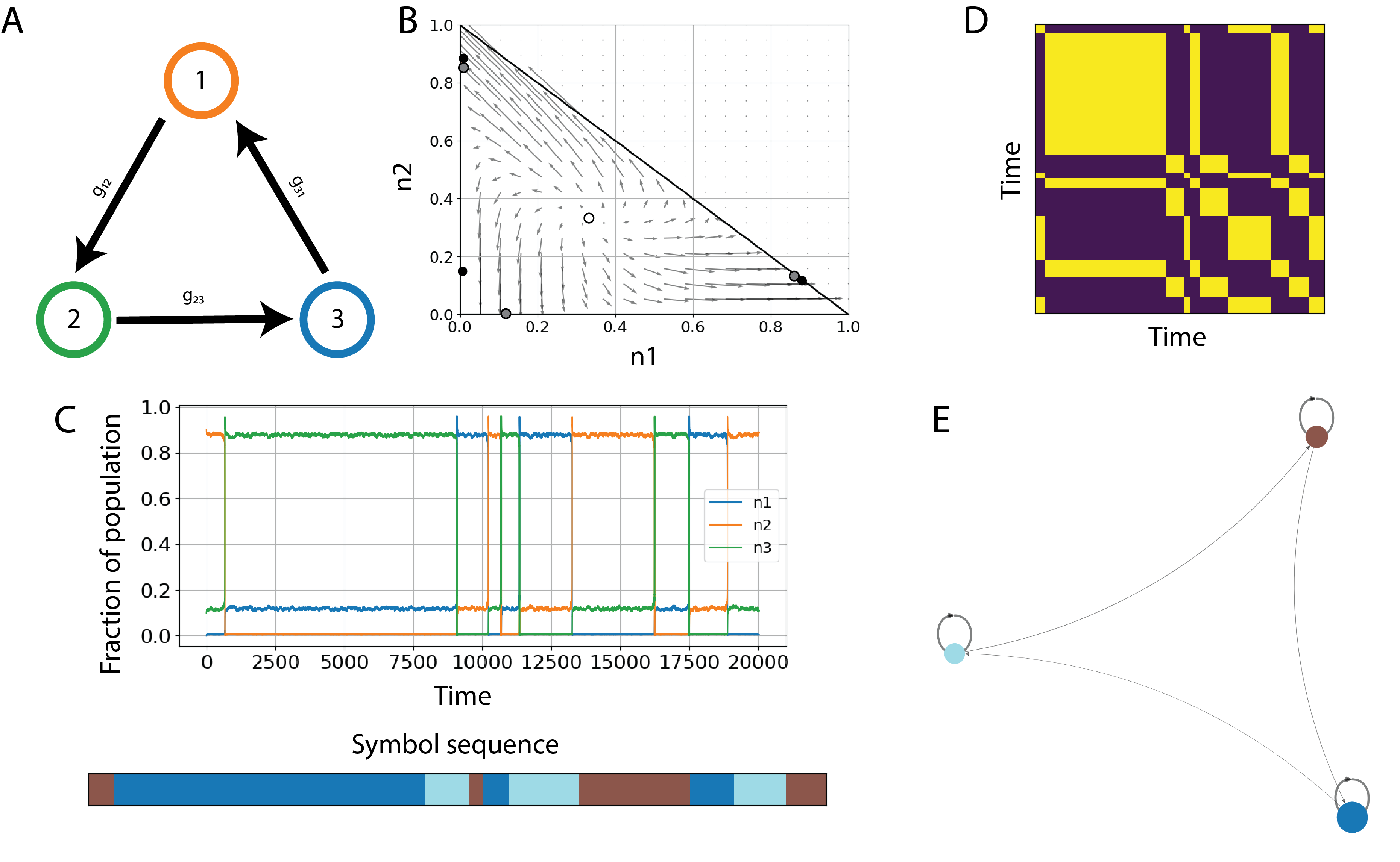}
\caption{Validation of state detection. To validate our pipeline we used a three-state stochastic oscillator model. The parameters used were $a = 3.105$, $N=15$, $U=1$, $V=0$, and $W=-1$. This combination of parameters corresponds to a phase where three symmetric metastable states emerge after a triple saddle-node bifurcation.
A) Schematic representation of a single unit's transitions between the three states (1, 2, 3) with stochastic transition rates $g_{12}$, $g_{23}$, and $g_31$ governed by the population distribution among the states.
B) Phase portrait showing the dynamics in the $n_1-n_2$ plane. The three black points represent the stable equilibria, the central white point indicates an unstable fixed point, and the gray points correspond to saddle points. The system exhibits disrupted limit cycles due to the bifurcation, leading to three center-saddle pairs.
C) Time series showing the fraction of the population in each state ($n_1$ in blue, $n_2$ in orange, and $n_3$ in green) over time. The sharp transitions between metastable states are driven by inherent fluctuations due to the stochastic nature of the model. Below the time series, a symbol sequence presenting the population's transitions between states.
D) Recurrence plot illustrating transitions over time. Alternating yellow and purple blocks indicate periods of stability followed by stochastic transitions between the metastable states.
E) Transition network derived from the time series, where nodes represent clusters of points corresponding to metastable states in the time series, and edges show the stochastic transitions between these regions of the phase space. The thickness of edges indicates transition frequency, while self-loops reflect persistence in metastable states.}
\label{toy_model_fig}
\end{figure*}

To validate the method presented in Fig. \ref{fig1}, we tested it using a model with three metastable states to demonstrate its ability to accurately capture transitions between different basins of attraction. Specifically, we aimed to show how the method can detect and characterize the topological features of system dynamics. Specifically, we selected a well-studied model of three-state stochastic oscillators \cite{wood2006universality}, where each unit (illustrated in Fig. \ref{toy_model_fig}A) transitions between three states based on probabilistic rules influenced by the number of neighboring units in each state. Each oscillator can be in one of the three states labeled 1, 2, and 3. At a given time point the oscillator can change its state from state $i$ to $j$ according to the transition rate $g_{ij}$ defined as:

\begin{equation}
g_{ij} = \exp \left[ a \left( U n_j + V n_{i-1} + W n_i \right) \right].
\end{equation}

For the specific combination of parameters $U = 1$, $V = 0$, and $W = -1$, and for a sufficiently large coupling parameter $a$, the mean-field model predicts three simultaneous saddle-node bifurcations, leading to the emergence of three symmetric stable states, as shown in Fig. \ref{toy_model_fig}B \cite{assis2011infinite}. When finite system size effects are considered, stochastic fluctuations become significant. These fluctuations induce instability in the previously stable states predicted by the mean-field theory, rendering them metastable and causing noise-induced transitions between these states.

To validate the pipeline shown in Fig. \ref{fig1}, we numerically simulated the Langevin dynamics of the model\cite{rosas2020synchronization}. The time series obtained from this simulation is presented in Fig. \ref{toy_model_fig}C. Using the values of $n_1$, $n_2$, and $n_3$, we computed a distance matrix for each point in the time series, performed a 3D embedding via Multidimensional Scaling, and then applied Gaussian Mixture Model clustering. The resulting symbolic sequence is shown at the bottom of Fig. \ref{toy_model_fig}C.

Three distinct symbols were identified for the time series, each representing a different metastable state. The three stable symbols correspond to the vicinity of the three stable states predicted by the mean-field approximation. These cyclic features are reflected in the recurrence plot in Fig. \ref{toy_model_fig}D. The quasi-diagonal pattern of continuous squares indicates the cyclic transitions between these symbols. The transition network in Fig. \ref{toy_model_fig}E effectively captures the topological characteristics of the cyclic transitions between the symbols. It clearly shows unidirectional transitions between the stable states, closely matching the topological features of the phase portrait in Fig. \ref{toy_model_fig}B.

\section{Symbolic Sequences of Joint Brain States}
\label{Symbolic_joint}

\begin{figure*}[htb]
\centering
\includegraphics[width=0.9\textwidth]{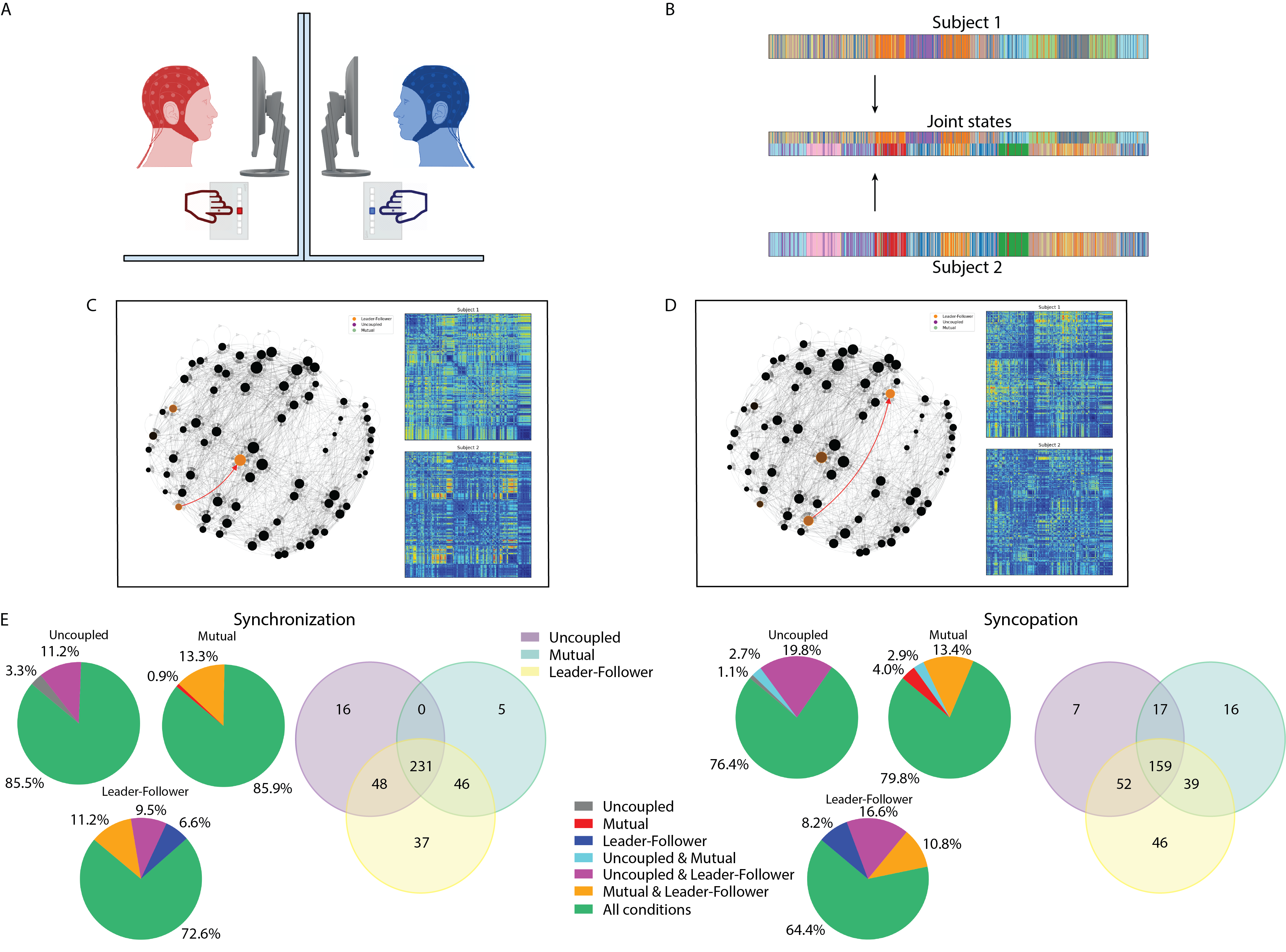}
\caption{A) Experimental setup used to investigate brain activity during motor coordination tasks with EEG. The experiment involved dyads, who performed a finger tapping task under different coordination patterns. In the Synchronization experiment, participants were instructed to synchronize their finger-tapping movements.  In the Syncopation experiment, participants aimed for movements intentionally out of phase (anti-phase synchronization). There were 12 trials in each experiment, each trial involving 150 taps at an average rate of 1.3 Hz. Three different interaction conditions were investigated: Uncoupled, where they were unaware of their partner's taps, Leader-Follower, where the Follower received feedback of the Leader's taps, and Mutual, where both individuals could receive feedback of each others taps.  B) Visualization of symbol sequences (represented by colors) for Subject 1 and Subject 2 in one experiment and their combined joint symbol sequence.  In this example, Subject 1 has 8 symbols, Subject 2 has 10 Symbols and the joint symbol sequence occupies 72 out of 80 of the possible joint symbols.  C) (Multimedia available online) Early dynamic evolution of the Joint symbol sequence can be represented as a trajectory in a transition network whose nodes are the possible joint states of the correlation matrices for each subject. The colored dots show the previous two states of the motif and the arrow shows the transition to the target node, with correlation matrices on the right. D) (Multimedia available online) Late dynamic evolution of the Joint symbol sequence can be represented as a trajectory in a transition network whose nodes are the possible joint states of the correlation matrices for each subject. The colored dots show the previous two states of the motif and the arrow shows the transition to the target node, with correlation matrices on the right.
E)Distribution of joint symbols across the experiments and trials.  The Venn diagram shows the overlap of unique joint symbols across the three interaction conditions. In both conditions the majority of joint symbols are observed in all 3 interaction conditions.  More interaction specific joint symbols are observed in syncopation than in synchronization.  The pie chart shows the fraction of observations of joint symbols, labeled by the interaction conditions in which they are observed. The majority of the time series remains in brain states that are represented in all interaction conditions.}  \label{fig2}
\end{figure*}

We can extend the notion of symbolic sequences to joint brain states when two (or more) individuals coordinate their behavior.  To demonstrate these ideas, we developed an experiment protocol to simultaneously record EEG from pairs of subjects (dyads) engaged in motor coordination patterns.  Each experiment involved a dyad performing a finger-tapping task aimed at maintaining a specific coordination pattern, as illustrated in Fig. \ref{fig2}A, either to \emph{synchronize} (in phase) or \emph{syncopate} (anti-phase) their finger-tapping movements. Three different interaction conditions were: \emph{Uncoupled}, where they were unaware of the other persons tapping, \emph{Leader-Follower}, where only one person received feedback of the other person tapping, and \emph{Mutual}, where both individuals could observe each others tapping. Further details of the experimental design are presented in Appendix \ref{AppendixExpt}. 
This experimental design allows us to precisely control the feedback conditions between participants. Feedback determines the joint behavior of the individuals in motor coordination\cite{moreau2023performance}, and we expected this manipulation would reorganize the dynamics of the joint brain states.  We hypothesize that variations in these feedback conditions will produce distinct effects on the stability and dynamical patterns of brain states, as reflected in symbolic sequences derived from EEG data. 
The EEG processing details are provided in Appendix \ref{AppendixEEG}. 

Brain states and symbol sequences were computed for each participant, and a joint symbol sequences were defined for each entire experiment, as the example shown in \ref{fig2}B. 
We model the joint symbol sequence through a transition network \cite{zou2019complex} in which each node represents a joint state and the links represents the transitions. \ref{fig2}C and \ref{fig2}D (Multimedia available online) shows a sequence of state changes (labeled by the arrows) within a representation of an entire session (all black nodes). On each frame the previous two node are represented by colored circles. The correlation matrices of each participant corresponding to the target node represented in orange are shown on the side of the figure. The transition network captures the evolution of joint brain states (functional connectivity) represented by the simultaneously observed correlation matrices.

A preliminary analysis of the symbol content for all the symbol sequences obtained leads to some interesting insights. If each participant has a time series represented as a sequence of $n_1$ and $n_2$ symbols respectively, the joint symbol sequence potentially has $n_1 \times n_2$ symbols distinct joint symbols, i.e., jointly observed correlation matrices. In the example show in Fig. \ref{fig2}, Subject 1 has 8 symbols, Subject 2 has 10 Symbols and the joint symbol sequence occupied 72 out of 80 possible joint symbols, across the entire experiment. In 7/12 experiments, every possible joint symbol was observed, and in another 5/12 more than 85 percent of the joint symbols were observed.  However, each individual trial only filled part of joint symbol space; in the vast majority (106/144) of the trials, fewer than 1/2 of the joint symbols were observed as shown in Fig. \ref{fig2}E. Most of the joint symbols were observed in all the interaction conditions.  Notably, synchronization exhibits more similarity in joint states across interaction conditions both in terms of the fraction of shared joint symbols (231/383 corresponding to 60\% for synchronization and 159/336 corresponding to 47\% for syncopation) and time spent in each joint symbol (for synchronization 70\% and for syncopation 60\% in states common in all three conditions).  Only a small fraction of time (10\% in synchronization, and 11\% in syncopation) was spent in states unique to each interaction condition. As discussed below, the distinction between interaction conditions was not in different brain states (joint symbols), but in the structure of symbolic dynamics.   
   
\section{Joint symbol recurrence quantification analysis}
\label{RQA}
\begin{figure*}[ht]
\centering
\includegraphics[width=0.9\textwidth]{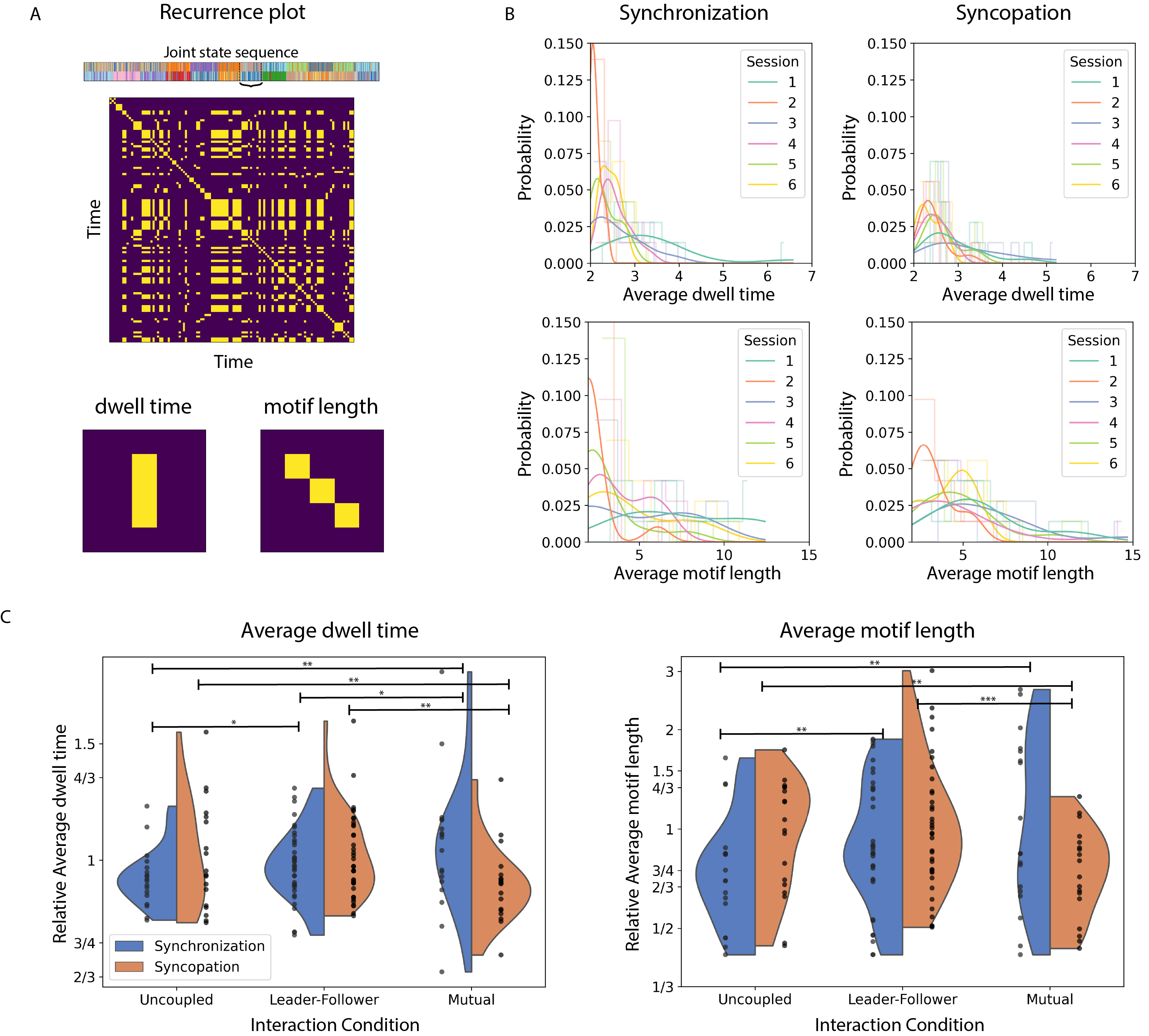}
\caption{A) Example of a recurrence plot of the joint state sequence over time. The top panel show the joint state sequence over time for the entire experiment with different colors representing different states.  The middle panel shows the recurrence plot for one trial (Mutual Interaction Condition) selected from the experiment.  The yellow pixels off diagonal in either vertical or horizontal indicate the repetition of the state on the diagonal.  The bottom panels depict examples of dwell time (left) and motif length (right) with value 3.
B) Distributions of average dwell time (top row) and average motif length O(bottom row) for synchronization (left column) and syncopation (right column) across six sessions. Each colored line represents a different experiment, showing how these metrics vary over different dyads.
C) Violin plots comparing the relative average dwell time (left) and relative average motif length (right) across different interaction conditions: Uncoupled, Leader-Follower, and Mutual. The blue violins represent synchronization sessions, and the orange violins represent syncopation sessions. The trial-averaged values were normalized across dyads, by dividing by the value of each trial by the average value for each experiment. The black dots show the values observed on individual trials. The black bars indicate statistics on the differences between conditions (*p < 0.1, **p < 0.05, ***p < 0.01). }\label{fig3}
\end{figure*}

We performed a recurrent quantification analysis (RQA) \cite{webber2015recurrence} specifically on the sequence of the dyad \emph{joint symbols}, where a joint symbol is a representation of simultaneous activity patterns across two individuals, to explore the temporal structure of the symbolic sequence throughout the entire session. Fig. \ref{fig3}A shows an example of the recurrence plot obtained for a single trial to identify dynamic patterns of the joint symbolic sequences. Here we explore two specific quantities, dwell time and motif length. The dwell time quantifies the size of continuous vertical lines in the recurrence plot, i.e., the number of consecutive time windows that the joint state of the two brains remained in the same joint symbol, while the motif length quantifies the length of diagonal lines (sequences of symbols) that are repeated during the experiment. The average dwell time and average motif length were calculated for each of the 12 trials in each experiment.

Fig. \ref{fig3}B, shows the distribution of trial average dwell time (upper panel) and trial average motif length (lower panel) for synchronization and syncopation sessions. Notably, both dwell time and motif length tend to be higher for syncopation compared to synchronization. The distributions also exhibit substantial variability between experiments, highlighting significant differences between dyads.  To examine differences in patterns across the three feedback regimes, we standardized these measures for each experiment, by dividing the trial average by the experiment average, creating what we refer to as the Relative scale of each measure. 

Figure \ref{fig3}C highlights that both average dwell time and average motif length are generally higher during syncopation sessions compared to synchronization sessions except for Mutual feedback condition. This indicates that the participants spent more time in specific states and formed longer sequences of repeated states (motifs) during syncopation, except for the Mutual feedback condition. We applied a bootstrap method to compare the mean of the distributions across different interaction conditions. 
Throughout the paper we define $\Delta\mu_{A,B}$ as the mean difference in a given measure (e.g., average dwell time or motif length) between two interaction conditions $A$ and $B$, where $A, B \in \{\text{U}, \text{LF}, \text{M}\}$ represent the Uncoupled, Leader-Follower, and Mutual conditions, respectively. Specifically, $\Delta\mu_{A,B} = \mu_A - \mu_B$, with negative values indicating that the mean for condition $A$ is lower than that for condition $B$.  
For synchronization coordination pattern, the mean difference in average dwell time between the Uncoupled and Leader-Follower conditions was $\Delta\mu_{U,LF} = -0.06$ ($p = 0.09$). The difference between the Uncoupled and Mutual conditions, $\Delta\mu_{U,M} = -0.15$ ($p = 0.04$), and the difference between Leader-Follower and Mutual conditions was $\Delta\mu_{LF,M} = -0.09$ ($p = 0.09$). Motif lengths were similarly modulated by interaction conditions in the synchronization task:  between the Uncoupled and Leader-Follower conditions, $\Delta\mu_{U,LF} = -0.30$ with a p-value of $p = 0.05$ and between the Uncoupled and Mutual conditions, $\Delta\mu_{U,M} = -0.45$ ($p = 0.03$). 
For the Syncopation experiment, neither dwell time nor motif length showed significant differences between the Uncoupled and Leader-Follower conditions. Dwell time in the Mutual condition was lower than Uncoupled, $\Delta\mu_{U,M} = 0.13$ ($p = 0.04$), and Leader-Follower and $\Delta\mu_{LF,M} = 0.14$ ($p = 0.01$).  Similarly, motif lengths were shorter in the Mutual interaction compared to Uncoupled, $\Delta\mu_{U,M} = 0.36$ ($p = 0.02$), and Leader-Follower interactions was $\Delta\mu_{LF,M} = 0.44$ ($p = 0.005$).

For synchronization, the Mutual interaction condition generally results in the highest average dwell time and motif length, indicating more stable and complex coordination patterns. Leader-Follower interactions showed intermediate dwell time and motif length.  In contrast, for syncopation, the Mutual condition shows the lowest average dwell time and motif length, suggesting less stable and less complex coordination patterns. These findings suggest that the type of interaction regime can significantly influence the complexity and stability of symbolic dynamics of brain networks.  

\section{Network topology analysis of the symbol sequence}
\label{Topology}

We used a transition network approach \cite{zou2019complex} to explore topological properties of the joint symbol sequences. We constructed a network representation of each trial in which each node represents a specific state of the subjects, with a node weight associated to the frequency of occurrence of the joint state. The edges between nodes are directed and indicate transitions between these states, and associated with each edge we defined a cost associated to the transition probabilities. To ensure the costs are additive, so that the cost for two consecutive transitions is the sum of the individual costs and reflects the probability of the sequence occurring, each edge's cost is defined as $-\ln(P)$, where $P$ is the transition probability.

\begin{figure*}[ht]
\centering
\includegraphics[width=0.9\textwidth]{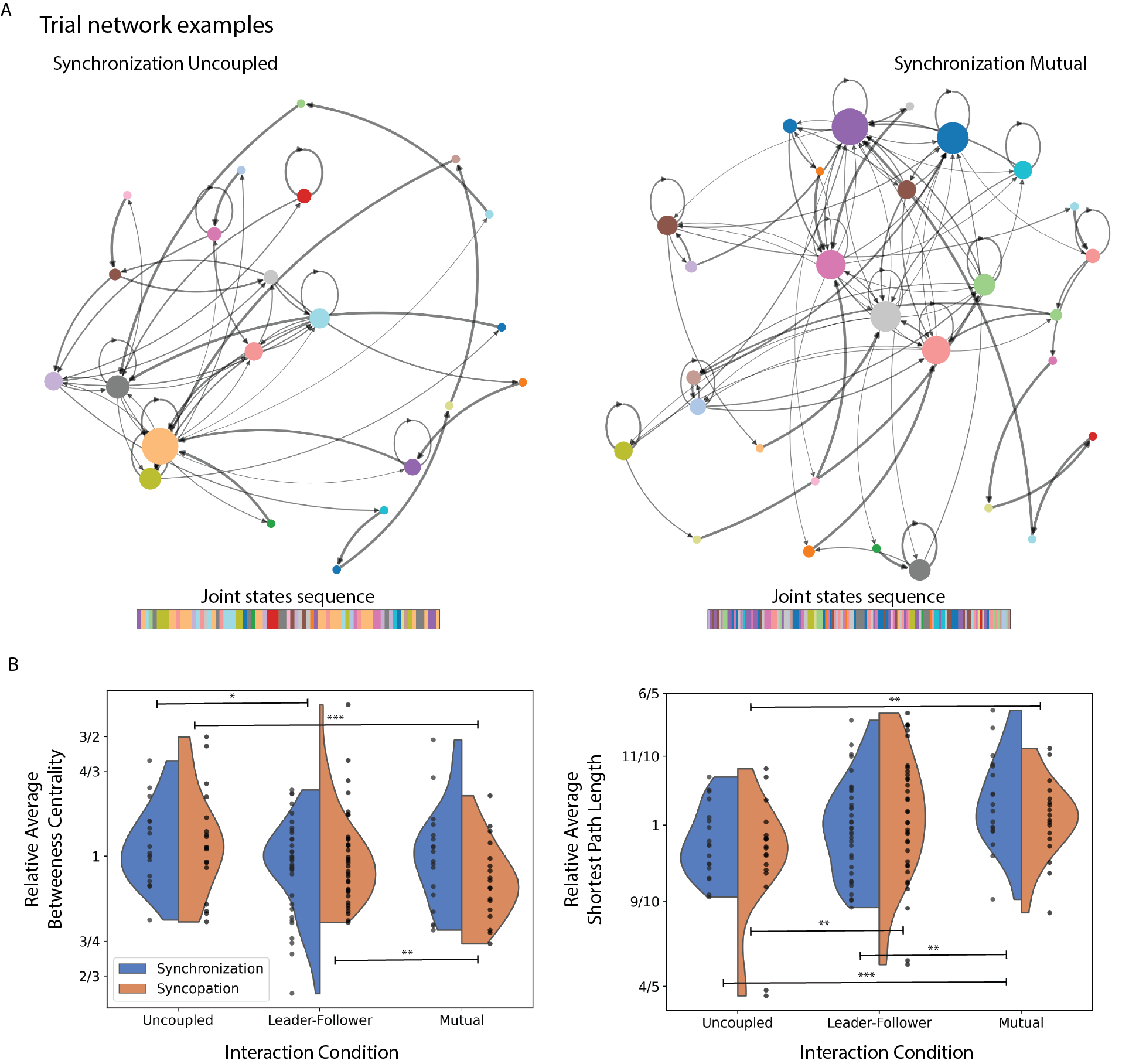}
\caption{A) Two examples of network representations: Uncoupled interaction on the left (average betweeness centrality = 0.13, average shortest path length = 3.30), and Mutual interaction on the right (average betweeness centrality = 0.08, average shortest path length = 4.16). The graphs summarizes the dynamic relationship between the dyad's joint symbols. Nodes represent joint symbols, with size proportional to their frequency. Edges indicate transitions, with a cost defined as as $-ln(P)$ based on the transition probabilities $P$, the edge width in the plot is proportional to its probability $P$. The horizontal bar below each graph depicts the sequence of joint states over time for this particular trial. This representation highlights the topology and dynamics of the observed joint states based on the state transitions.
B) Violin plots comparing the relative average betweenness centrality (left), and relative average shortest path length (right) across different feedback conditions: Uncoupled, Leader-Follower, and Mutual. The blue violins represent synchronization sessions, and the orange violins represent syncopation sessions. The black bars indicate significant differences between conditions (*p < 0.1, **p < 0.05, ***p < 0.01).}\label{fig4}
\end{figure*}

Figure \ref{fig4}A presents examples of such network representations. In this plot, the node distances reflect the distances estimated from the correlation matrices shown in Figure \ref{fig1}B. The joint sequence for this trial is presented at the bottom, with colors matching the nodes. The size of the nodes is proportional to their weight, and the thickness of the edges is proportional to the transition probability represented by the edge.

In our network topology analysis we found that the shortest path length and  betweenness centrality plays an important role in differentiating the 3 interaction conditions. With this network representation, the shortest path between two nodes represents the sequence of state transitions with the highest overall probability, as each edge's cost is defined as $-\ln(P)$, where $P$ is the transition probability. This means the shortest path highlights the most probable route in terms of transition probabilities from one state to another. Betweenness centrality measures the extent to which a specific state serves as a crucial intermediary in these high-probability sequences. A state with relatively high betweenness centrality, compared to other states in the system, frequently appears in the most probable paths connecting other states. This indicates its importance as a core node that facilitates the flow of dynamics throughout the system.representing a state that is part of the several paths linking other states, making it crucial for the flow of dynamics in the system. The average shortest path length reflects the efficiency of the network's state transitions, while the average betweenness centrality indicates the degree of centralization and the potential for bottlenecks.

To compare across dyads, we normalized the measures obtained for each trial against the average value for each experiment.  The results in Fig. \ref{fig4}B show that for synchronization tasks, there are no significant differences in the  Average Betweenness Centrality values. However, for  Average Shortest Path Length, significant differences were found between Uncoupled and Mutual ($\Delta\mu_{U,M} = -0.079$, p=0.004) and Leader-Follower and Mutual ($\Delta\mu_{LF,M} = -0.057$, p=0.022), with the Mutual condition always demonstrating longer path lengths. This indicates a lower spread of probability flow along more states, but since the Average Betweenness Centrality does not change, it suggests that the amount of flow through a core structure in the network is maintained.

For syncopation tasks, Relative Average Betweenness Centrality was significantly lower in the Mutual condition compared to both Uncoupled ($\Delta\mu_{U,M} = 0.187$, p=0.0085) and Leader-Follower ($\Delta\mu_{LF,M} = 0.109$, p=0.042), indicating less centralized probability flow. Additionally, significant differences in Shortest Path Length were found between Uncoupled and Leader-Follower ($\Delta\mu_{U,LF} = -0.077$, p=0.017) and Uncoupled and Mutual ($\Delta\mu_{U,M} = -0.074$, p=0.016), suggesting a lower spread of states in the presence of either type of interaction condition. However, unlike the synchronization sessions, this reduction of the spread of the flow reduces the importance of the core states as indicated by the lower Betweenness Centrality. 

\section{Discussion}
\label{Discussion}
Our approach to defining symbols departs from other approaches \cite{beim2016optimal} by employing Gaussian Mixture Models (GMM) to identify the symbols by focusing on regions with a high density of points. A particular characteristic of GMM is its capability of modeling these dense regions as a mixture of Gaussian distributions, capturing the underlying probability density effectively. This probabilistic approach allows GMM to handle clusters with varying shapes and sizes and provides a likelihood to evaluate the model using Bayesian Information Criterion (BIC). Therefore, GMM provides an appropriate method for identifying these high-density regions which can be associated with basins of attraction.
The differences between clustering methods, particularly for the definition of recurrence plots can be a potential topic for future investigations on methods to detect basins of attraction from time series.

During dyadic motor coordination tasks, symbolic dynamics reveals that the type of interaction conditions significantly influences the average dwell time of the observed joint states (vertical lines in the recurrence matrix) and the average motif length (diagonal lines in the recurrence matrix). This indicates that the feedback regime affects the stability of the joint brain states and the complexity of the dynamical patterns of the joint states. Importantly, the interaction conditions patterns lead to opposite effects in the two tasks: Mutual feedback results in longer average dwelling times and average motif lengths during synchronization, while during syncopation, these quantities are shorter. These results suggest that different mechanisms play a role in the different coordination patterns. 

To explore the topological features of the states, we conducted a network-based analysis of the symbolic dynamics. We defined transition networks based on the symbol sequence for each trial, to identify different topological changes as the feedback regimes are implemented in each coordination pattern. In the synchronization condition, the results showed that stronger coupling led to an increase in average dwell time and motif length, while betweenness centrality remained constant and average shortest path length increased. These findings suggest that stronger coupling stabilizes a few states, making the probability flow more restricted and preserving the core-periphery structure of the network. 
A strong core-periphery structure implies the system has a set of dominant, highly interconnected basins of attraction (the core) where transitions are frequent and stable. The peripheral basins of attraction are less frequently visited, with transitions mainly leading back to the core. This structure indicates that the system tends to remain within the core's stable dynamics and only occasionally explores the less stable, transient states represented by the periphery.  In contrast, in the syncopation condition, the average dwelling time and motif length decreased, betweenness centrality declined, and average shortest path length increased. This indicates that stronger coupling in syncopation enhances the stability of a larger set of states and distributes the probability flow more broadly, reducing the dominance of core nodes and creating a more distributed but less efficient flow.

To investigate the relationship between the number of states, RQA measures, and transition network measures, we performed correlation analyses and examined scatter plots of each pair of measures. We expected to find correlations, particularly between dwell time, betweenness centrality, and motif length. As anticipated, the results showed clear correlations between these measures. Specifically, the recurrence plot measures were correlated due to the contribution of central blocks of long dwell times to both vertical and diagonal lines. We found a correlation between dwell time and motif length $r=0.66$ for Synchronization sessions and $r = 0.71$ for Syncopation. Similarly, the correlation between RQA measures and betweenness centrality was expected, as higher betweenness centrality reflects greater stability of the states. We found a correlation of betweenness centrality and dwell time $r = 0.51$ for Synchronization and $r = 0.59$ for Syncopation. These findings are consistent with the expected, as the association between state stability and both recurrence and network measures is expected.

\section{Conclusion}
\label{Conclusion}

We investigated brain activity as a series of transitions between metastable states to analyze multi-brain symbolic sequences. Our results and analytical approach expand the concept of brain metastable states to multi-person interactions, illustrating how the landscape of these states changes under different experimental conditions. Utilizing a data-driven approach, we defined brain states based on the dissimilarity of correlation matrices between different brain areas for each subject, allowing us to create symbolic representations of joint states. We then analyzed these symbolic sequences using recurrence quantification analysis and a transition network approach. Recurrence quantification analysis showed that feedback regimes significantly influenced the duration at which the dyad remains in a joint brain state, and the length of repeated sequences of joint brain states. A network-based analysis of symbolic dynamics revealed different topological changes with feedback regimes and task. These findings provide insights into how feedback between individuals affects state stability and dynamics in motor coordination tasks.

Our work offers a fresh perspective on inter-brain connectivity by analyzing a hyperscanning neuroimaging dataset through multi-brain symbolic dynamics. Metastable states dynamics are crucial for brain flexibility, enabling seamless transitions between stable functional states and supporting cognitive processes and resilience to noise through robust mechanisms\cite{rabinovich2020sequential,tognoli2014metastable,kelso2006metastability,roberts2019metastable}.   Our study expands the concept of brain metastable states to multi-person interactions, providing a novel perspective without relying on millisecond time-scale synchronization. Our results and analytical approach not only broaden the understanding of brain metastable states in the context of human interaction and coordination but also open new avenues for studying the brain mechanisms underlying these processes. This new tool bridges brain dynamics and behavior, offering a powerful and flexible method to explore the intricate neural choreography of multi-person interactions in a broad range of cognitive functions and collaborative behaviors.

\begin{acknowledgments}
This research was sponsored by the U.S. Army DEVCOM Army Research Laboratory and was completed under Cooperative Agreement Number W911NF2420013 and by the National Science Foundation grant 2126976. The views and conclusions contained in this document are those of the authors and should not be interpreted as representing the official policies, either expressed or implied, of the U.S. Army DEVCOM Army Research Laboratory or the U.S. Government. The U.S. Government is authorized to reproduce and distribute reprints for Government purposes notwithstanding any copyright notation herein.
\end{acknowledgments}

\section*{Data Availability Statement}

The Python code used in the analysis and the symbolic sequences obtained can be found in \url{https://github.com/italoivo/Motorcoordination_Symbolic_Dynamics}
The EEG data that support the findings of this study are available from the corresponding author upon reasonable request.

\appendix
\section{Experimental Protocol}
\label{AppendixExpt}

Six pairs of subjects (from a pool of 11 total, 4 M and 7 F), each performed one session of synchronized tapping and one session of syncopated tapping with each other on separate days. Each participant was visually and sound isolated from the other participant.

There were three interaction conditions in each session:
\begin{enumerate}
    \item \emph{Uncoupled} In the synchronized tapping session: the dyads tapped independently on their own following a sequence of 30-tap pacing stimulus (1.3Hz). They each received visual feedback of their own individual taps (with a self-feedback signal (colored dot) on the screen).  In the syncopated tapping session: the dyads attempted to tap in between visual feedback of a  tapping sequence generated by themselves during the previous synchronization session. Thus, they are syncopating with a random signal generated independently from their own tapping history in the synchronized tapping session. 
    \item \emph{Leader-Follower} In the synchronized tapping session, one subject (Leader) received the 30-tap pacing stimulus, then tapped independently (self feedback), in the same way as in the Uncoupled condition. The other subject (Follower) tried to synchronize with real-time visual feedback of the tapping sequence by the Leader. In the syncopated tapping session, one subject (Leader) attempted to tap in between the tapping sequence, generated by themselves during the synchronization session,  in the same way as in the Uncoupled condition. The other subject (Follower) tried to syncopate with the real-time visual feedback of the tapping sequence of the Leader. Both individuals in a dyad were placed in the role of Leader or Follower in separate trials.
    \item \emph{Mutual} In the synchronized tapping session: Following synchronization with a 30-tap pacing stimulus at 1.3 Hz, each participant in the dyad tried to synchronize with the other participant, while receiving real-time visual feedback of the other participant’s tapping. In the syncopated tapping session: Following syncopation with a 30-tap pacing stimulus at 1.3 Hz,  each participant in the dyad tried to syncopate with the other participant, while receiving real-time visual feedback of the other participant’s tapping.
\end{enumerate}
In each experiment session, there were 12 trials, 3 Uncoupled, 6 Leader-Follower with roles exchanged, and 3 Mutual presented in a random order.  Each trial terminated when one of the participants reached 230 taps.
\section{EEG data acquisition and processing}
\label{AppendixEEG}

The following subsections describe the EEG acquisition and processing steps performed before the symbolic representation was obtained.

\subsection{EEG data acquisition and preprocessing}
EEG data was acquired from the dyads using a pair of identical 32-channel TMSi EEG system. The sampling rate was set to 2000 Hz. Raw EEG data were band-pass filtered between 0.25 and 50 Hz and underwent ICA (Independent component analysis). Independent components highly correlated with EEG data from channels Fp1 and Fp2 were marked as eye movement artifacts for removal.

\subsection{Source localization procedure, forward and inverse solution}

The locations of EEG sensors were spatially aligned with the digitized scalp landmarks using “fiducials” from the standardized Freesurfer fsaverage surfaces with MNE’s coregistration method.

Source space was created using a parameter of icosahedron subdivision grade 4.
The conductivity of the three layers for EEG is 0.3, 0.0075, and 0.3. Boundary element method (BEM) was used to create a forward model of the head geometry using the standardized fsaverage surface. This procedure created 5124 source locations with free orientations, which were then converted to surface-based source orientations with each dipole having a fixed orientation perpendicular to the surface. It produced a leadfield matrix with the size of 32 sensors by 5124 dipoles. 

We applied an inverse solution on the cleaned scalp EEG by inverting the reduced lead field using regularized minimum norm estimation (weighted L2 norm) \cite{dale1993improved}.
PCA to aggregate sources within each ROI:
We then used PCA (Principal component analysis) to aggregate the source-localized data of 5124 dipoles for each one of the 448 cortical brain regions. The 448 cortical regions are based on scale 250 of the Lausanne Parcellation \cite{daducci2012connectome}.

\subsection{EEG data processing}
After the transformation to source space, the EEG signal sequentially filtered with the two Butterworth filters using a zero-phase filtering method: a low-pass filter with a cutoff frequency of 50 Hz and 20 dB loss in the stop band, designed to remove higher-frequency noise from the signal, and a high-pass filter with a cutoff frequency of 0.5 Hz and 20 dB loss in the stop band and a designed to remove low frequency artifacts. The filtered signal is then downsampled by a factor of 10, reducing the sampling rate to 200 Hz. Next, the Hilbert transform is applied to to obtain its analytical representation. Both the real and imaginary parts of the analytical signal are smoothed using the Savitzky-Golay filter, which helps reduce noise while preserving the signal's features. The smoothed complex signal is segmented into overlapping 2s time windows with a 50\% overlap. 

\section{Metric properties of the logarithm of the inverse cosine similarity}

\subsection*{Correlation matrices definition and properties}

The entries of the complex correlation matrix $r_{ij}$ are defined as the complex Pearson correlation \cite{vsverko2022complex} between the complex-valued signals $x_i$ and $x_j$:

\[
r_{ij} = \frac{\sum_{n=1}^N (x_{i,n} - \langle x_i \rangle)(x_{j,n}^* - 
    \langle x_j^* \rangle)}{\sqrt{\sum_{n=1}^N |x_{i,n} - \langle x_i \rangle|^2 \sum_{n=1}^N |x_{j,n} - \langle x_j \rangle|^2}}
\]

Observe that this definition implies that the entries $r_{ij}$ can be interpreted as inner products, this fact implies that the complex correlation matrices are Gram matrices, from which stems the hermitian property $r_{ij} = r_{ji}^*$ and positive semi-definite (real non-negative eigenvalues).

Now, consider the correlation matrices $A$ and $B$. Then we can write their spectral decompositions as:
\[
A = U \Lambda_A U^\dagger \quad \text{and} \quad B = V \Lambda_B V^\dagger
\]
where \( U \) and \( V \) are unitary matrices, and \( \Lambda_A \) and \( \Lambda_B \) are diagonal matrices with non-negative eigenvalues \( \lambda_i \) and \( \mu_j \), respectively.

Using the decompositions of \( A \) and \( B \), we have:
\[
\operatorname{Tr}(A B) = \operatorname{Tr}(U \Lambda_A U^\dagger V \Lambda_B V^\dagger).
\]
By the cyclic property of the trace, this becomes:
\[
\operatorname{Tr}(A B) = \operatorname{Tr}(\Lambda_A C \Lambda_B C^\dagger),
\]
where \( C = U^\dagger V \) is also unitary.

Expanding \( \operatorname{Tr}(\Lambda_A C \Lambda_B C^\dagger) \) in terms of matrix elements, we get:
\[
\operatorname{Tr}(\Lambda_A C \Lambda_B C^\dagger) = \sum_{i,j} \lambda_i |C_{ij}|^2 \mu_j.
\]
Since \( \lambda_i \geq 0 \), \( \mu_j \geq 0 \), and \( |C_{ij}|^2 \geq 0 \), each term in the sum is non-negative. Thus,
\[
\operatorname{Tr}(A B) \geq 0.
\]
Which guarantees that the trace between two correlation matrices $A$ and $B$ are real and non-negative

\subsection*{Correlation matrices distance}

We define the distance between correlation matrices \(A\) and \(B\) as:
\[
d(A, B) = \log \left(\frac{\|A\| \|B\|}{\text{Tr}(A^{\dagger} B)}\right)
\]
where \(\|A\|\) is the Frobenius norm of matrix \(A\), and \(\text{Tr}(A^{\dagger} B)\) is the trace of the matrix product $A^{\dagger} B$. Observe that the argument of the logarithm is the complex version of the widely used cosine similarity measure, and for the correlation matrices are guaranteed to be real and non-negative.  

We show that this function satisfies three conditions of a metric: non-negativity, symmetry, and identity of indiscernibles. However, the triangle inequality is not guaranteed to hold for all cases of correlation matrices. To address this, we provide a counterexample and discuss the specific cases in which the triangle inequality fails for this metric.

\subsection*{Non-negativity}
To show \(d(A, B) \geq 0\), observe that the argument of the logarithm is real and positive as \(\|A\|_F > 0\), \(\|B\|_F > 0\), and \(\text{Tr}(A^{\dagger} B) \geq 0\). Furthermore, \(\text{Tr}(A^{\dagger} B) \leq \|A\|_F \|B\|_F\) by the Cauchy-Schwarz inequality, so \(d(A, B) \geq 0\).

\subsection*{Symmetry}
Using the cyclic property of the trace:
\[
\text{Tr}(A^{\dagger} B) = \text{Tr}(B^{\dagger} A)
\]
we have:
\[
d(A, B) = \ln\left(\frac{\|A\|_F \|B\|_F}{\text{Tr}(A^{\dagger} B)}\right) = d(B, A)
\]

\subsection*{Identity of Indiscernibles}
If \(A = B\), then \(\text{Tr}(A^{\dagger} A) = \|A\|_F^2\), so:
\[
d(A, A) = \ln\left(\frac{\|A\|_F^2}{\|A\|_F^2}\right) = \ln(1) = 0
\]
If \(d(A, B) = 0\), then by the Cauchy-Schwarz inequality, \(\text{Tr}(A^{\dagger} B) = \|A\|_F \|B\|_F\), implying \(A = B\).

\subsection*{Triangle Inequality}
We aim to explore the triangle inequality:
\[
d(A, C) \leq d(A, B) + d(B, C)
\]
By properties of logarithms, we can rewrite the inequality as:
\[
\ln\left(\frac{\|A\|_F \|C\|_F}{\text{Tr}(A^{\dagger} C)}\right) \leq \ln\left(\frac{\|A\|_F \|B\|_F}{\text{Tr}(A^{\dagger} B)}\right) + \ln\left(\frac{\|B\|_F \|C\|_F}{\text{Tr}(B^{\dagger} C)}\right)
\]
which simplifies to:
\[
\frac{\|A\|_F \|C\|_F}{\text{Tr}(A^{\dagger} C)} \leq \frac{\|A\|_F \|B\|_F}{\text{Tr}(A^{\dagger} B)} \cdot \frac{\|B\|_F \|C\|_F}{\text{Tr}(B^{\dagger} C)}
\]
Now, consider the following:
\[
\frac{\|A\|_F \|C\|_F}{\text{Tr}(A^{\dagger} C)} \leq \frac{\|A\|_F \|B\|_F}{\text{Tr}(A^{\dagger} B)} \cdot \frac{\|B\|_F \|C\|_F}{\text{Tr}(B^{\dagger} C)}
\]
Multiplying the terms gives:
\[
\frac{\|A\|_F \|C\|_F}{\text{Tr}(A^{\dagger} C)} \leq \frac{\|A\|_F \|C\|_F}{\text{Tr}(A^{\dagger} C)} \cdot \frac{\|B\|_F^2 \text{Tr}(A^{\dagger} C)}{\text{Tr}(A^{\dagger} B) \cdot \text{Tr}(B^{\dagger} C)}
\]
From this result, we have that the triangle inequality holds for
\[
\frac{\|B\|_F^2 \text{Tr}(A^{\dagger} C)}{\text{Tr}(A^{\dagger} B) \cdot \text{Tr}(B^{\dagger} C)} \geq 1
\]

To explore this inequality, we ran several numerical tests and found that the above inequality holds for a wide range of correlation matrices, including all the matrices in our dataset. However, we also identified a counterexample that violates the triangle inequality.

Consider a configuration involving three specific matrices, $ A $, $ B $, and $ C $, chosen to reveal this limitation. Matrix $ A $ is a correlation matrix with all off-diagonal entries equal to $-r$, where $ r $ is constrained by $ 0 < r \leq \frac{1}{n - 1} $ to maintain positive semi-definiteness. This setup represents maximal anticorrelation among the variables. Matrix $ B $ is the identity matrix $ I $, while matrix $ C $ is fully correlated, with all entries equal to 1, depicting maximal positive correlation. For this configuration, we observe that $\text{Tr}(A^{\dagger} C) = 0$, which violates the inequality. Notably, this case relates to the scenario in which $ d \to \infty $, indicating that for very large distances, the triangle inequality does not hold.

Despite this theoretical limitation, extensive numerical tests conducted on EEG datasets have demonstrated that the proposed distance performs effectively in practical applications. One contributing factor to this effectiveness is the inherent limit on the anticorrelation coefficients when analyzing many variables. For an $n \times n$ matrix it is $\frac{1}{n - 1}$. This theoretical limit guarantees the non-negativity of the correlation matrix eigenvalues; Physically it means that many signals cannot be strongly mutually anticorrelated. Recent studies have also analyzed the triangle inequality for cosine similarity \cite{schubert2021triangle}.

\bibliography{aipsamp}

\end{document}